\documentclass[amsmath,amssymb,aps,pre,superscriptaddress,nofootinbib,notitlepage]{revtex4-1}

\usepackage{graphicx}
\usepackage{mathtools,color}

\newcommand{\dd}{\mathrm{d}}

\begin{document}

\title{Relaxation Processes in Long-Range Lattices}

\author{T.~M.~Rocha Filho}
\affiliation{Instituto de F\'\i sica and International Center for Condensed Matter Physics\\ Universidade de
Bras\'\i{}lia, CP: 04455, 70919-970 - Bras\'\i{}lia, Brazil}

\author{R.~Bachelard}
\affiliation{Departamento de F\'{\i}sica, Universidade Federal de S\~{a}o Carlos, Rod. Washington Lu\'{\i}s, km 235 - SP-310, 13565-905 S\~{a}o Carlos, SP, Brazil}

\begin{abstract}
The relaxation to equilibrium of lattice systems with long-range interactions is investigated. The timescales involved depend polynomially on the system size, potentially leading to diverging equilibration times. A kinetic equation for long-range lattices is proposed, which explain these timescales as well as a threshold in the interaction range reported in [Phys.\ Rev.\ Lett.\ {\bf 110}, 170603 (2013)]. Non-Markovian effects are shown to play an important role in the relaxation of systems of up to thousands of particles.
\end{abstract}

\maketitle

\section{Introduction}

Systems with long-range interactions are present at various scales in nature, from astrophysics to atomic scales~\cite{Campa2009}. Apart from
the peculiarities in their statistical equilibrium, such as negative specific heat~\cite{lyndenbell1977} or ensemble
inequivalence~\cite{barre2001}, an important feature of these systems are the evolution time scales that arise.
For example, governed by the gravitational force, galactic dynamics is characterized by its inefficiency at redistributing kinetic energy~\cite{binney},
resulting in relaxation times which increase with the system size. At the microscopic scale, long-range effects rise in the presence of light-mediated
(dipole-dipole) interactions between cold atoms~\cite{Chalony2013,Barre2014,Aspelmeyer2014}, and size-dependent equilibration times were also
reported~\cite{guerin2016}.

Long-range interactions may manifest between either the internal or external degrees of freedom of the particles, which leads to different scaling laws
for the relaxation times~\cite{Gabrielli2010}. In this context, a new platform for the study of long-range lattices, i.~e.\ for particles at fixed
positions but with dynamical internal variables, has emerged with the realization of ion chains with tunable interactions~\cite{britton2012,senko2015}.
There, the coupling between the particles decay as an adjustable power-law of their mutual
distance, which makes ion traps a versatile platform to study the crossover from short- to long-range, including in the quantum realm~\cite{richerme2014,jurcevic2014}.

From a theoretical point of view, explicit scaling laws were derived for the a few specific interactions: Astrophysical systems present relaxation times that scale as $N/\log N$, with $N$ the number of interacting bodies; For the so-called Hamiltonian mean-field model~\cite{antoni1995},
with infinite-range interactions, a scaling with $N$ for non-homogeneous and $N^2$ for homogeneous states~\cite{yamaguchi2003,bouchet2005,rochafilho2014} was derived.
The existence of size-dependent time scales was reported in the more general context of long-range lattices
with $1/r^\alpha$ interactions ($\alpha$ smaller than $d$ the dimension of the system for the long-range case),
both in the classical and in the quantum context~\cite{kastner2011,bachelard2013}. In particular, a puzzling threshold in the scaling laws was reported at $\alpha=d/2$, which does not correspond to the transition from long-range to short-range systems.

In this work we derive explicitly the scaling of the relaxation time of one-dimensional ($d=1$) long-range classical lattices. The threshold at $\alpha=d/2$ originates in two-particle correlation terms, and it manifests both in the non-Markovian (small size) and Markovian (large size) regimes. Indeed, the derivation of a kinetic equation for long-range lattices allows to identify the specific scaling of the contribution of non-Markovian terms, which are all the more important for fast-decaying interactions terms ($\alpha\to d$).

\section{Kinetic equation for long-range lattices}

\subsection{Long-range lattices}

Let us consider a one-dimensional lattice of $N$ particles at coordinates $x_k=k\xi$, $k=1,..N$ and $\xi$ the lattice step. Each particle has an internal degree of freedom represented by an angular coordinate $\theta_k$ and its canonically conjugate momentum $p_k$, with a Hamiltonian of the form:
\begin{equation}
H(\mathbf{p},\mathbf{\theta})=\sum_{k=1}^N\frac{p_k^2}{2}+\frac{1}{2\tilde N}\sum_{\substack{j,k=1\\(j\neq k)}}^N\frac{v(\theta_j-\theta_k)}{r_{jk}^\alpha},
\label{eq1}
\end{equation}
where $v(\theta_j-\theta_k)/r_{jk}^\alpha$ is the pair interaction potential for particles $j$ and $k$, separated by the distance $r_{jk}=\| x_j-x_k\|$, with $v(0)=0$. The factor $1/\tilde N\equiv 1/N^{1-\alpha}$ in front of the potential energy term is introduced
(e.~g.\ by a renormalization of time) in order to obtain an extensive total energy and to properly define the passage to a Vlasov equation description.
Such interaction was investigated as a generalization of the Hamiltonian Mean-Field model~\cite{anteneodo1998}, and describes more generally lattices
of fixed particles interacting through their internal degrees of freedom, as for example in light scattering problems~\cite{Lehmberg1970,maximo2015,hill2017}.


The evolution of the system and its relaxation are studied introducing the $N$-particle distribution function $f_N(1,2,\ldots,N;t)$, defined such that $f_N\, d1 \cdots dN$ is the probability for particle $k$ ($k=1,2,\ldots N$) to be in the volume element $dk\equiv d\theta_kdp_k$ that contains the point $(\theta_k,p_k)$ in phase-space, at time $t$.
We note that since all particles are localized on the lattice points, the distribution $f_N$ is not required to be invariant with respect to particle permutations,
which must be taken into account in the determination of the generalized form of the BBGKY hierarchy.

The Liouville equations for the lattice system then writes as
\begin{eqnarray}
\frac{\partial f_N}{\partial t} & = & \left\{H,f_N\right\} = \sum_{k=1}^N\left\{\frac{\partial H}{\partial\theta_k}\frac{\partial f_N}{\partial p_k}
-\frac{\partial H}{\partial p_k}\frac{\partial f_N}{\partial\theta_k}\right\}
\nonumber\\
 & = & \sum_{k=1}^N\left\{-\frac{\partial f_N}{\partial\theta_k}p_k+\frac{1}{2\tilde N}\sum_{\substack{l=1\\(l\neq k)}}^N\frac{1}{r_{kl}^\alpha}\frac{\partial v_{kl}}{\partial\theta_k}
\frac{\partial f_N}{\partial p_k}\right\},
\label{eq2}
\end{eqnarray}
with $v_{kl}\equiv v(\theta_k-\theta_l)$.

\subsection{Generalized BBGKY hierarchy}
Since each particle is distinguished by its position on the lattice, in order to define the $s$-particle reduced distribution function we have to specify
which particle positions and momenta are integrated out. Thus,
the $s$-particle reduced distribution function depending on particles of indices $i_1,\ldots,i_s$ is defined by:
\begin{equation}
f_s(i_1,\ldots,i_s)\equiv\int \prod_{j\neq i_1,\ldots,i_s}\dd j\:f_N(1,\ldots,N).
\label{eq3}
\end{equation}
In Eq.~(\ref{eq3}) and from now on we keep the time dependence implicit, except where necessary.
For the one-particle distribution function ($s=1$), the specification on which particle phase-space coordinates it depends is not required, provided boundary
effects are negligible. Note that these reduced distribution functions are not invariant by permutation of their arguments in a lattice, in general.

Let us now consider two disjoint sets of different particle indices $J_1$ and $J_2$, such that $J_1$ has $s$ indices and $J_2$ has $N-s$ indices.
By integrating Eq.~(\ref{eq2}) over the position and  momentum variables of particles in $J_2$ we obtain:
\begin{eqnarray}
\frac{\partial}{\partial t}f_s(J_1)&=&\frac{\partial}{\partial t}\int\prod_{j\in J_2}\dd j\:f_N(1,\ldots,N)
\nonumber\\
 & = &\int\prod_{j\in J_2}\dd j\:\left\{-\sum_{k=1}^N\frac{\partial f_N}{\partial\theta_k}p_k
+\frac{1}{2\tilde N}\sum_{\substack{k,l=1\\(k\neq l)}}^N\frac{1}{r_{kl}^\alpha}\frac{\partial v_{kl}}{\partial\theta_k}\partial_{kl}f_N\right\},
\label{eq4}
\end{eqnarray}
where we introduced the cross derivative $\partial_{jk}\equiv\partial/\partial p_j-\partial/\partial p_k$.
By eliminating vanishing surface terms in Eq.~(\ref{eq4}) and using Eq.~(\ref{eq3}) we  obtain the generalized form of the BBGKY hierarchy:
\begin{eqnarray}
\frac{\partial}{\partial t}f_s(J_1)&=&-\sum_{k\in J_1}p_k\frac{\partial}{\partial\theta_k}f_s(J_1)
+\frac{1}{2\tilde N}\sum_{\substack{k,l\in J_1\\(k\neq l)}}\frac{1}{r_{kl}^\alpha}\frac{\partial v_{kl}}{\partial\theta_k}\partial_{kl}f_s(J_1)
\nonumber \\
 & & +\frac{1}{\tilde N}\sum_{k\in J_1}\sum_{\substack{l=1\\ (l\neq k)}}^N
\int\dd l\:\frac{1}{r_{kl}^\alpha}\frac{\partial v_{kl}}{\partial\theta_k}
\frac{\partial}{\partial p_k}f_{s+1}(J_1^{(l)}),
\label{eq5}
\end{eqnarray}
where $J_1^{(l)}\equiv J_1\cup\{l\}$.
The next step towards the kinetic equation is the introduction of the $s$-particle correlation functions, as discussed below.

\subsection{Irreducible cluster representation and prototypical kinetic equation}

The $s$-particle reduced distribution function can be decomposed into a purely uncorrelated part and contributions from
$s$-particle correlation functions $g_s(j_1,\ldots,j_s)$ as~\cite{balescu1997}:
\begin{eqnarray}
 f_2(j,k) &= &f_1(j)f_1(k)+g_2(j,k),\nonumber \\
 f_3(j,k,l)&= &f_1(j)f_1(k)f_1(l)+\sum_{P(j,k,l)}f_1(j)g_2(k,l)+g_3(j,k,l),\label{eq6}
\end{eqnarray}
and so on, where $P(j,k,l)$ stands for all different permutation of particles $j,k,l$. All permutations must be considered due to the absence of permutation invariance.
Since we are considering correlation among different particles, in the sum in Eq.~(\ref{eq6}) only terms with $k\neq l$ must be considered for
$g_2(k,l)$.
In order to determine the order of magnitude of the functions $g_s$, we note that the correlation between two particles require the interaction
between these particles, the correlation function $g_2$ is of order $\tilde N^{-1}$,
the order of the interaction. Similarly, $g_3$ requires at least two pair interactions, and is therefore of order $\tilde N^{-2}$.
More generally, the correlation function $g_s$ is of order $\tilde N^{-s+1}$.

The kinetic equation is a closed equation for the
one-particle reduced function, and therefore as a preliminary step we replace Eq.~(\ref{eq6}) into Eq.~(\ref{eq5}) for $s=1$ ($J_1={j}$) to obtain:
\begin{widetext}
\begin{equation}
\frac{\partial}{\partial t}f_1(j)  = -p_j\frac{\partial}{\partial\theta_j}f_1(j)+\frac{1}{\tilde N}
\sum_{\substack{k=1\\ (k\neq j)}}^N\int\dd k\frac{1}{r_{jk}^\alpha}
\frac{\partial v_{jk}}{\partial\theta_j}\frac{\partial}{\partial p_j}\left[f_1(j)f_1(k)+g_2(j,k)\right].\label{eq7}
\end{equation}
Proceeding similarly for $s=2$ (with $J_1=\{j,k\}$) leads to:
\begin{eqnarray}
\frac{\partial}{\partial t}f_2(j,k) &=&-p_j\frac{\partial}{\partial\theta_j}f_2(j,k)
-p_k\frac{\partial}{\partial\theta_k}f_2(j,k)
+\frac{1}{\tilde N}\frac{1}{r^\alpha_{jk}}\frac{\partial v_{jk}}{\partial\theta_k}\partial_{jk}f_2(j,k)
\nonumber\\
 & & +\frac{1}{\tilde N}\sum_{\substack{l=1\\ (l\neq j)}}^N\int\dd l\:
\frac{1}{r_{jl}^\alpha}\frac{\partial v_{jl}}{\partial\theta_j}\frac{\partial}{\partial p_j}f_3(j,k,l)
     +\frac{1}{\tilde N}\sum_{\substack{l=1\\ (l\neq k)}}^N\int\dd l\:\frac{1}{r_{kl}^\alpha}
\frac{\partial v_{kl}}{\partial\theta_k}\frac{\partial}{\partial p_k}f_3(j,k,l).
\label{eq8}
\end{eqnarray}
Hence, in order to have a closed equation for $f_1$ the correlation function $g_2$ must be expressed in terms of $f_1$.
Replacing the irreducible cluster expansion in Eq.~(\ref{eq6}) into Eq.~(\ref{eq8}) yields:
\begin{eqnarray}
&& \left(\frac{\partial}{\partial t}+ p_j\frac{\partial}{\partial\theta_j}+p_k\frac{\partial}{\partial\theta_k}\right)g_2(j,k)=
\frac{1}{\tilde N}\:\frac{1}{r_{jk}^\alpha}\frac{\partial v_{jk}}{\partial\theta_j}\partial_{jk}\left[f_1(j)f_1(k)+g_2(j,k)\right]
\nonumber \\ && +\frac{1}{\tilde N}\sum_{\substack{l=1\\ (l\neq j)}}^N\int\dd l\:\frac{1}{r_{jl}^\alpha}\frac{\partial v_{jl}}{\partial\theta_j}\frac{\partial}{\partial p_j}
\left[f_1(j)g_2(k,l)+f_1(k)g_2(j,l)\right]
 + \frac{1}{\tilde N}\sum_{\substack{l=1\\ (l\neq k)}}^N\int\dd l\:\frac{1}{r_{kl}^\alpha}\frac{\partial v_{kl}}{\partial\theta_k}\frac{\partial}{\partial p_k}
\left[f_1(j)g_2(k,l)+f_1(k)g_2(j,l)\right]
\nonumber\\
 & + &
\frac{1}{\tilde N}\sum_{\substack{l=1\\ (l\neq k)}}^N\int\dd l\:\frac{1}{r_{kl}^\alpha}\frac{\partial v_{kl}}{\partial\theta_k}\frac{\partial}{\partial p_k}
g_3(j,k,l)
+\frac{1}{\tilde N}\sum_{\substack{l=1\\ (l\neq j)}}^N\int\dd l\:\frac{1}{r_{jl}^\alpha}\frac{\partial v_{jl}}{\partial\theta_j}\frac{\partial}{\partial p_j}
g_3(j,k,l),
\label{eq9}
\end{eqnarray}
\end{widetext}
where we used $\int_{-\pi}^\pi\dd\theta_l (\partial v_{kl}/\partial\theta_l)=0$.

By considering only terms up to order $1/\tilde N$, one obtains a closed form equation for $g_2$. Its solution,
inserted in Eq.~(\ref{eq7}), and after some simplifying assumptions (see below) leads to the generalization of the Balescu-Lenard equation~\cite{balescu1997}.
It is known that, in the mean-field case ($\alpha=0$), the collision term in the kinetic equation vanishes at order $1/\tilde N$ for a
homogeneous state~\cite{sano2012}. Consequently, one must consider terms of order $1/\tilde N^2$ by including
three-particle correlations to describe the relaxation process. Nevertheless, solving the
resulting equation for $g_3$ is a daunting task. Here we rather follow the weak-coupling approach of
Ref.~\cite{rochafilho2014}: it consists in expanding the hierarchy in orders of a weak coupling, considering an inter-particle potential of order $\lambda\ll1$, but at the same time
retaining only dominant terms in $1/\tilde N$. In the case of long-range systems, where each particle is coupled to all others,
it refers to disordered regimes, where the macroscopic degrees of freedom that describe the coupling (such as the magnetization)
are very weak on average. For a detailed discussion of the weak
coupling approach, the reader is referred to Ref.~\cite{rochafilho2014,Fouvry2019}. This leads to
a generalization of the Landau equation for lattice systems.
The correlations functions are then expanded as:
\begin{eqnarray}
 & & g_2(j,k)=\lambda g^{(1)}_2(j,k)+\lambda^2 g^{(2)}_2(j,k)+{\cal O}\left(\lambda^3\right),
\nonumber\\
 & & g_3(j,k,l)=\lambda^2 g^{(2)}_3(j,k,l)+{\cal O}\left(\lambda^3\right).
\label{corrlmbexp}
\end{eqnarray}
Before determining a kinetic equation at leading order in $\lambda$ and $1/\tilde N$, let us first consider the mean-field limit for the lattice system.

\subsection{Vlasov equation}

The mean field description is obtained in the limit $\tilde N\rightarrow\infty$ 
and is equivalent to neglecting two-particle correlations $g_2={\cal O}(\tilde N^{-1})$
in Eq.~(\ref{eq7}). It results in the generalized form of the Vlasov equation~\cite{Bachelard2011}:
\begin{eqnarray}
\frac{\partial}{\partial t}f_1(j)&=&-p_j\frac{\partial}{\partial\theta_j}f_1(j)
\label{eq9b}
\\ &&+\frac{1}{\tilde N}\sum_{\substack{k=1\\ (k\neq j)}}^N\int\dd k\frac{1}{r_{jk}^\alpha}
\frac{\partial v_{jk}}{\partial\theta_j}\frac{\partial}{\partial p_j}f_1(j)f_1(k),\nonumber
\end{eqnarray}
after eliminating a vanishing surface term. The continuous limit on the lattice is obtained by performing $d\rightarrow0$
at constant lattice length ${\cal N}=Nd$:
\begin{eqnarray}
\frac{\partial}{\partial t}&&f_1(\theta,p)=-p_j\frac{\partial}{\partial\theta}f_1(\theta,p)\label{eq10}
\\ && +\frac{1}{\cal N}
\int\dd x^\prime\dd\theta^\prime\dd p^\prime\:\frac{1}{|x^\prime|^\alpha}
\frac{\partial v(\theta-\theta^\prime)}{\partial\theta}\frac{\partial}{\partial p}f_1(\theta,p)f_1(\theta^\prime,p^\prime).
\nonumber
\end{eqnarray}
In Eq.~(\ref{eq10}) collisional (granularity) effects are neglected. For states homogeneous in $\theta$, one has $\partial f_1(p)/\partial t=0$,
i.~e., the distribution function has no dynamics. An evolution is recovered by introducing the corrections due to collisions, which is the purpose of the next sections. The less trivial case of non--homogeneous
states can be addressed using action-angle variables~\cite{chavanis2007}, yet it will not be considered here.

\subsection{Weak coupling limit}

For the homogeneous states under consideration, the prototypical kinetic equation in Eq.~(\ref{eq7}) assumes the form:
\begin{equation}
\frac{\partial}{\partial t}f_1(j)=\frac{1}{\tilde N}
\sum_{\substack{k=1\\ (k\neq j)}}^N\frac{1}{r_{jk}^\alpha}\frac{\partial}{\partial p_j}
\int\dd k\frac{\partial v_{jk}}{\partial\theta_j}g_2(j,k).
\label{eq7b}
\end{equation}
We now proceed to obtained a closed form for the two-particle correlation function $g_2$ in the weak-coupling limit
by performing the expansion in power series on the parameter $\lambda$ given in Eq.~(\ref{corrlmbexp}).
This represents a generalization of the Landau equation for lattice systems.

By replacing Eq.~(\ref{corrlmbexp}) into Eq.~(\ref{eq9}) and retaining only dominant terms in $\lambda$ we obtain:
\begin{equation}
\left(\frac{\partial}{\partial t}+p_j\frac{\partial}{\partial\theta_j}+p_k\frac{\partial}{\partial\theta_k}\right)g^{(1)}_2(j,k)=
\frac{1}{\tilde N}\:\frac{1}{r_{jk}^\alpha}\frac{\partial v_{jk}}{\partial\theta_j}\partial_{jk}f_1(j)f_1(k).
\label{eq10b}
\end{equation}
As discussed in Ref.~\cite{rochafilho2014}, this approximation is justified by the fact that the total effective force
on a given particle 
is weak in a non-magnetized state. Equation~(\ref{eq10b}) can be solved in the form of a convolution as:
\begin{eqnarray}
\lefteqn{g^{(1)}_2(j,k;t)=e^{(-p_j\partial/\partial\theta_j-p_k\partial/\partial\theta_k)t}g^{(1)}_2(j,k;0)}
\label{eq11}\\
 & & +\frac{1}{\tilde N}\int_0^t\dd\tau\:e^{(-p_j\partial/\partial\theta_j-p_k\partial/\partial\theta_k)\tau}
\frac{1}{r_{jk}^\alpha}\frac{\partial v_{jk}}{\partial\theta_j}\partial_{jk}f_1(j;t-\tau)f_1(k;t-\tau).
\nonumber
\end{eqnarray}
The first term on the right-hand side of this equation is a transient contribution from the correlations at the initial time,
and can be discarded after a short transient~\cite{balescu1997}. Since for a homogeneous
state, the angle variables $\theta_j$ evolve in a free (ballistic) motion, up to corrections of order $\lambda$, we obtain
\begin{equation}
f_1(j;t-\tau)f_1(k;t-\tau)=e^{(p_j\partial/\partial\theta_j+p_k\partial/\partial\theta_k)\tau}f_1(j;t)f_1(k;t)
	+{\cal O}\left(\lambda\right),
\label{eq11b}
\end{equation}
where we have used the free time propagator $\exp(-t\:\partial/\partial\theta)$. Using the identities
\begin{eqnarray}
e^{(-p_j\partial/\partial\theta_j-p_k\partial/\partial\theta_k)\tau}&\frac{\partial}{\partial p_j}&
e^{(p_j\partial/\partial\theta_j+p_k\partial/\partial\theta_k)\tau}
=\frac{\partial}{\partial p_j}+\tau\frac{\partial}{\partial\theta_j},
\nonumber
\\ e^{(-p_j\partial/\partial\theta_j-p_k\partial/\partial\theta_k)\tau}&\frac{\partial}{\partial\theta_j}&v(\theta_j-\theta_k)=
\frac{\partial}{\partial\theta_j}v(\theta_{jk}-p_{jk}\tau),
\label{eq11d}
\end{eqnarray}
where $\theta_{jk}\equiv\theta_j-\theta_k$ and $p_{jk}\equiv p_j-p_k$, we rewrite equation~(\ref{eq11}) as
\begin{equation}
g^{(1)}_2(j,k;t)=\frac{1}{\tilde N}\int_0^t\dd\tau\:
\frac{1}{r_{jk}^\alpha}v^\prime(\theta_{jk}-p_{jk}\tau)\partial_{jk}f_1(j;t-\tau)f_1(k;t-\tau).
\label{eq12}
\end{equation}
In particular, the latter equation implies that
\begin{equation}
g^{(1)}_2(j,k)=g^{(1)}_2(k,j).
\label{g2antisim}
\end{equation}
Plugging Eq.~(\ref{eq12}) into Eq.~(\ref{eq7b}), and using that fact that the mean-field force cancels for a homogeneous state, we obtain:
\begin{widetext}
\begin{equation}
\frac{\partial}{\partial t}f_1(p_j;t)=\frac{\lambda}{\tilde N^2}\sum_{\substack{k=1\\(k\neq j)}}^N\frac{1}{r_{jk}^{2\alpha}}\int_0^t\dd\tau\:
\int\dd p_k\:\partial_{jk}\:\langle{\cal F}_{jk}(0){\cal F}_{jk}(\tau)\rangle\:
\partial_{jk}f_1(p_j;t-\tau)f_1(p_k;t-\tau),
\label{eq12b}
\end{equation}
where the force auto-correlation function has been introduced:
\begin{equation}
\langle{\cal F}_{jk}(0){\cal F}_{jk}(\tau)\rangle=\int_{-\pi}^\pi\dd\theta_k\frac{\partial}{\partial\theta_j}v(\theta_{jk})
\frac{\partial}{\partial\theta_j}v(\theta_{jk}-p_{jk}\tau).
\label{forcecorr}
\end{equation}
\end{widetext}
A kinetic equation in the form of Eq.~(\ref{eq12b}) is clearly non-Markovian, and one usually goes a step further:
if the force auto-correlation decays to zero faster than any significant change in the one-particle
distribution function, then the time integral in Eq.~\eqref{eq12b} can be extended to infinity and we can set $f_1(p;t-\tau)\rightarrow f_1(p;t)$.
This results in a Markovian dynamical evolution~\cite{balescu1997,lourenco2015}.
The validity of this approximation is discussed in details in Sec.~\ref{sec:Markov}.

Let us first show that, assuming the Markovian regime is reached, the first-order contribution to the kinetic equation is dominant. To this end, expressions~(\ref{eq12b}) and~(\ref{forcecorr}) are cast in the Fourier space, using the following series for the potential  $v(\theta)$:
\begin{equation}
v(\theta)=\sum_{n=-\infty}^\infty \tilde v(n)e^{-in\theta},
\label{eq12c}
\end{equation}
where the Fourier coefficients are given by:
\begin{equation}
\tilde v(n)=\frac{1}{2\pi}\int_{-\pi}^\pi\dd\theta\:v(\theta)e^{in\theta},
\label{eq12d}
\end{equation}
and the correlation function by:
\begin{equation}
\tilde g_2^{(l)}(n,m; p_i,p_j; t)=\int_{-\pi}^\pi\dd\theta_i\int_{-\pi}^\pi\dd\theta_j
\:g_2^{(l)}(i, j ;t)\:e^{in\theta_i}e^{im\theta_j}.
\label{corrfunexp}
\end{equation}
From Eq.~(\ref{eq12}) we obtain:
\begin{eqnarray}
\tilde g^{(1)}_2(n, m; p_j,p_k;t)&=&-\frac{in\delta_{n+m,0}}{(2\pi)^2r_{jk}^\alpha}\:\tilde v(n)\partial_{jk}f_1(p_j;t)f_1(p_k;t)\nonumber
\\ &&\times \int_0^\tau\dd\tau\:e^{inp_{jk}\tau},
\label{eq12e}
\end{eqnarray}
where $\delta_{n,m}$ refers to the Kronecker delta.
As $g^{(1)}_2(j,k)$ is the correlation function between two different particles, it is only defined for $j\neq k$.
Equation~\eqref{eq7b} then converts into:
\begin{widetext}
\begin{eqnarray}
\frac{\partial}{\partial t}f_1(p_j;t) & = & \frac{2\pi\lambda}{\tilde N}\sum_{\substack{l=1\\(l\neq j)}}^N\frac{1}{r_{jl}^{2\alpha}}\int_{-\infty}^\infty\dd p_l
\sum_{n=-\infty}^\infty\left[n\tilde v(n)\right]^2\partial_{jl}\int_0^\infty\dd\tau\: e^{inp_{jl}\tau}\partial_{jl}f_1(p_j;t)f_1(p_l;t),
\nonumber\\
 & = & \frac{2\pi^2\lambda}{\tilde N}\sum_{\substack{l=1\\(l\neq j)}}^N\frac{1}{r_{jl}^{2\alpha}}\sum_{n=-\infty}^\infty\left|n\tilde v(n)\right|^2\partial_{jl}
\delta_+(np_{jl})\partial_{jl}f_1(p_j;t)f_1(p_l;t),
\label{eq13}
\end{eqnarray}
\end{widetext}
where we have used the property $v(-n)=v^*(n)=v(n)$ and the Cauchy integral:
\begin{equation}
\int_0^\infty\dd\tau\:e^{ia\tau}=\pi\delta_+(a)=\pi\delta(a)+i{\cal P}\left(\frac{1}{a}\right),
\label{eq13b}
\end{equation}
with ${\cal P}(x)$ the Cauchy principal part of $x$ (an odd function). The imaginary part of the right-hand side of Eq.~\eqref{eq13}
vanishes and setting $p_j=p$ and $p_l=p^\prime$, we obtain:
\begin{eqnarray}
 & & \frac{\partial}{\partial t}f_1(p;t)=\frac{2\pi^2\lambda}{\tilde N}\sum_{\substack{l=1\\(l\neq j)}}^N\frac{1}{|x_j-x_l|^{2\alpha}}
\sum_{n=-\infty}^\infty\left|n\tilde v(n)\right|^2\label{eq14}\\
 & & \times\frac{\partial}{\partial p}\int_{-\infty}^\infty\dd p^\prime\:
\delta(n(p-p^\prime))\left(\frac{\partial}{\partial p}-\frac{\partial}{\partial p^\prime}\right)f_1(p;t)f_1(p^\prime;t).\nonumber
\end{eqnarray}
The right-hand side of Eq.~(\ref{eq14}) vanishes due to the Dirac delta function, just as for the Balescu-Lenard
equation for the mean-field ($\alpha=0$) case~\cite{rochafilho2014}.
Thus, under the hypothesis of a Markovian dynamics, one needs to go one order further in $\lambda$ to determine the kinetic equation.

\subsection{Higher-order contributions\label{sec:higherorder}}

Neglecting the two last terms in the right-hand side of Eq.~(\ref{eq9}), i.~e., retaining only terms up to order $1/\tilde N$ and neglecting collective effects, leads to an integral equation for $g_2$. Its solution, inserted into Eq.~(\ref{eq7b}), will provide a generalization of the Balescu-Lenard equation for lattice systems~\cite{balescu1997,lenard1960}. Once more, it results in a vanishing collisional integral.
Indeed, the only contributions to $g_2$ with a non-vanishing contribution to the kinetic equation originate in the three-particle correlation function $g_3$ of Eq.~(\ref{eq7b}), which are of order $1/\tilde N^2$.

Using algebra package MAPLE~\cite{maple}, the leading order contribution for $g_3$ are determined to be
\begin{widetext}
\begin{eqnarray}
	\left[\frac{\partial}{\partial t}
	-p_k\frac{\partial}{\partial\theta_k}-p_l\frac{\partial}{\partial\theta_l}-p_n\frac{\partial}{\partial\theta_n}\right]
	g_3(k,l,n;t)=
-\sum_{m=1}^N\left[\frac{D_k^{(l,n,m)}}{r_{k,m}^\alpha}+\frac{D_l^{(k,n,m)}}{r_{l,m}^\alpha}+
	 \frac{D_n^{(k,l,m)}}{r_{n,m}^\alpha}\right]
	 -\frac{B_{k,l}^{(n)}}{r_{k,l}^\alpha}-\frac{B_{k,n}^{(l)}}{r_{k,n}^\alpha}-\frac{B_{l,n}^{(k)}}{r_{l,n}^\alpha},
	\label{eqC32}
\end{eqnarray}
where we have introduced
\begin{eqnarray}
	B_{k,l}^{(n)}&=& v^\prime(\theta_k-\theta_l)\partial_{kl}
\left[f(p_k,t)g_2^{(1)}(\theta_l,\theta_n,p_l,p_n,t)+f(p_l,t)g_2^{(1)}(\theta_k,\theta_n,p_k,p_n,t)
	+f(p_n,t)g_2^{(1)}(\theta_k,\theta_l,p_k,p_l,t)\right],
	\label{defBs}
\\ 		D_k^{(l,n,m)}&=&\frac{\partial}{\partial p_k}f(p_k,t)v^\prime(\theta_k-\theta_m)
	\left[
		f(p_l,t)\int\dd\theta_m\dd p_m \:g_2^{(1)}(\theta_n,\theta_m,p_n,p_m,t)
		+f(p_n,t)\int\dd\theta_m\dd p_m \:g_2^{(1)}(\theta_l,\theta_m,p_l,p_m,t)\right].\nonumber
\end{eqnarray}
\end{widetext}
Inserting the three-particle correlation $g_3$ from Eq.~\ref{eqC32} into Eq.~\eqref{eq9}, one can obtain a closed form for
the kinetic equation, to the second order in the coupling. This approach was used in Ref.~\cite{rochafilho2014} to obtain
a kinetic equation for the Hamiltonian mean-field model.
Since we are here interested in the {\it scaling} of the relaxation time,
rather than its exact expression, we now proceed with discussing the Markov approximation.


\section{Relaxation times and the Markov approximation\label{sec:Markov}}

\subsection{The Markovian hypothesis}

The Markov approximation consists in assuming that the force auto-correlation function $C_p$ in Eq.~(\ref{forcecorr}) vanishes over a time
such that the one-particle distribution function does not change significantly. If it does not apply, auto-correlation terms in the force distribution makes that the right-hand term of Eq.~\eqref{eq12b} contributes substantially, and may even dominate over the higher-order contributions discussed in Sec.~\ref{sec:higherorder}.

The (non-)Markovian nature of the dynamics strongly depends on the model, especially on the system size~\cite{balescu1997}, but also on the initial conditions.
Let us here consider the cosine potential from the $\alpha$XY chain~\cite{anteneodo1998,bachelard2013}:
\begin{equation}
	v(\theta)=-\cos(\theta),
	\label{vpotdef}
\end{equation}
for which we study numerically the evolution of the force auto-correlation. The simulations are realized by integration of the equation of motion, where the initial state is a random realization of a state homogeneous in angles and bounded in momentum:
\begin{equation}
        f(p,\theta,t=0)=
        \left\{
                \begin{array}{l}
                        1/(4\pi p_0),\:\:{\rm if}\:\:-p_0<p<p_0;
                        \\
                        0,\:\:{\rm otherwise}.
                \end{array}
                \right.
        \label{wbstate}
\end{equation}
The constant $p_0$ allows to choose an energy such that the system remains in a non-magnetized phase at all times~\cite{Bachelard2011}.

\begin{figure}[ht]
\centering
\includegraphics[width=0.8\textwidth]{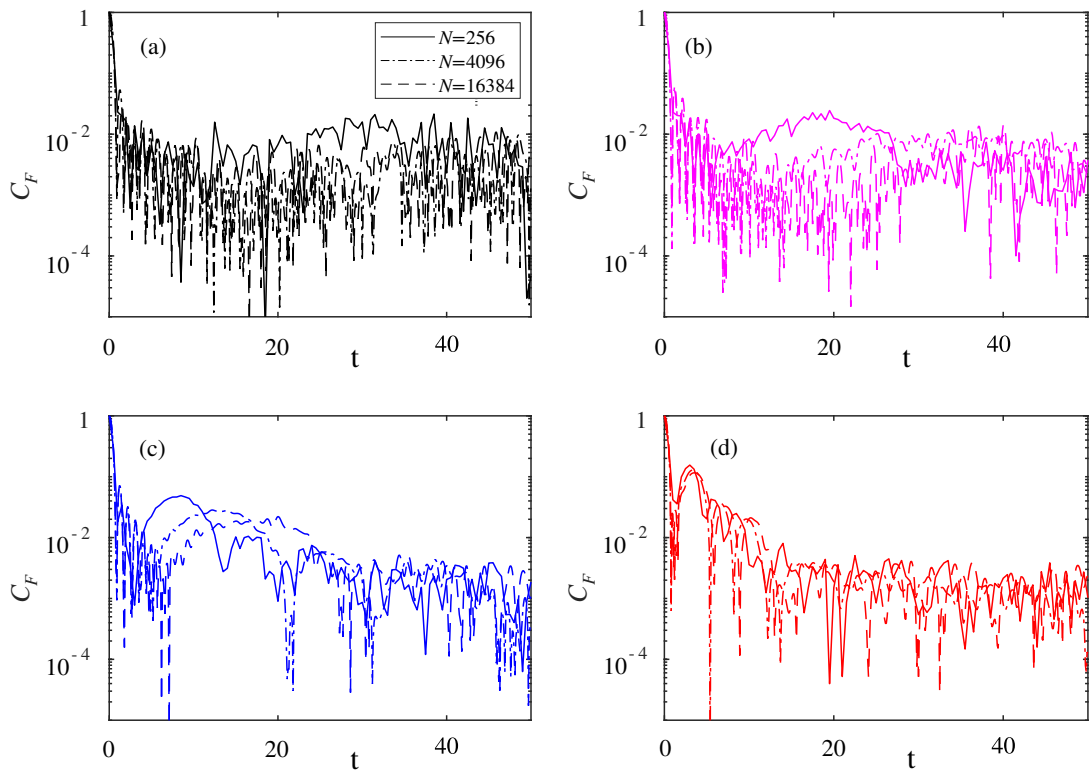}
\caption{Evolution of the auto-correlation of the force for different system sizes and interaction ranges $\alpha$. The plain lines correspond to $N=256$, the dash-dotted ones to $N=4096$ and the dashed ones to $N=16384$ for a) $\alpha=0$, b) $\alpha=0.3$, c) $\alpha=0.6$ and d) $\alpha=0.9$. Simulations realized for $p_0=7$.}
	\label{Fig:CF}
\end{figure}

The Markovian nature of the dynamics is tested by comparing the dynamics of the average auto-correlation function of the force $C_F$ with the macroscopic evolution of the distribution function,
and is assessed from molecular dynamics simulations. 
This average auto-correlation function $C_F$ is defined by:
\begin{equation}
	C_F(\tau)\equiv \frac{\sum_{j,k\neq j}\langle{\cal F}_{jk}(0){\cal F}_{jk}(\tau)\rangle}{\sum_{j,k\neq j}\langle{\cal F}_{jk}(0){\cal F}_{jk}(0)\rangle},
\label{forcecorr2}
\end{equation}
so it is normalized at time zero.
Here we implemented the numerical solutions of the Hamiltonian
equations of motion using a fourth-order symplectic integrator~\cite{yoshida1990} in a parallel implementation as described in Ref.~\cite{filho2014}.
Because we are considering homogeneous states, we monitor the first moments of the momentum distribution $M_k\equiv\langle p^k\rangle$.
The odd moments fluctuate around zero, and the second moment $M_2$ does not evolve substantially due to energy conservation.
We thus focus on $M_4$ (other even momentum $M_6$, $M_8$, $\ldots$ support the same conclusions but higher flutuations with higher order),
which typically present a slow dynamics during the relaxation process~\cite{bachelard2013}.
Figure~\ref{Fig:CF} shows the force auto-correlation as a function of time for an increasing system size $N$
(ranging from $256$ to $16384$) and several values of $\alpha$.
The decay of the force auto-correlation does not appear to depend significantly on the system size, and little
on the interaction range $\alpha$. We thus consider that it cancels around $t_{FC}\approx 30$, and consider
that the Markovian approximation is valid if the fourth momentum $M_4$ does not vary significantly in this time interval,
i.~e.\ that $M(t_{FC})-M_4(0)$ is small.

\begin{figure}[ht]
\centering
\includegraphics[width=0.8\textwidth]{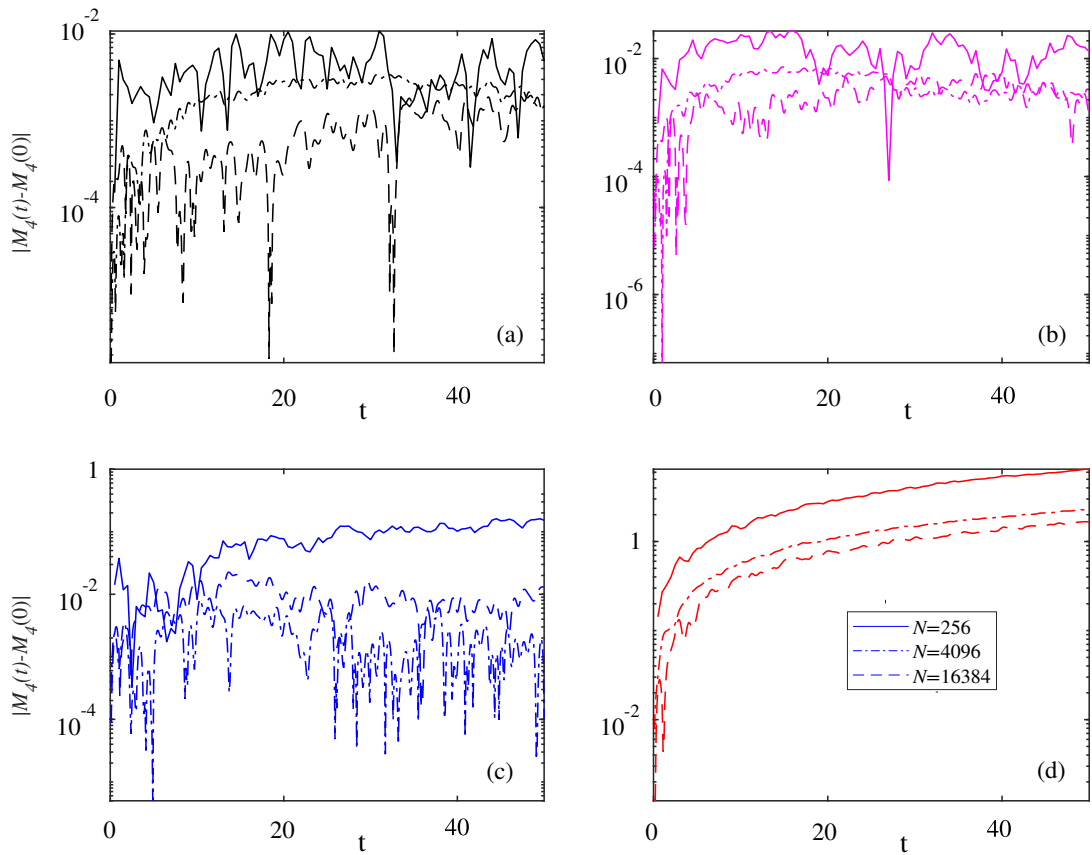} 
\caption{Evolution of the fourth momentum $M_4$ for different system sizes and interaction ranges $\alpha$
        for a) $\alpha=0$, b) $\alpha=0.3$, c) $\alpha=0.6$ and d) $\alpha=0.9$.
        The initial value the momentum is always close to $M_4(0)\approx30$. The plain lines correspond to $N=256$,
        the dash-dotted ones to $N=4096$ and the dashed ones to $N=16384$. Simulations realized for $p_0=7$.}
	\label{Fig:M4}
\end{figure}

Differently from the force auto-correlation, the momentum $M_4$ presents an evolution that depends strongly on both the system size and the interaction range.
Figure~\ref{Fig:M4} shows the variation of $M_4$ from its initial value as a function of time.
Larger $N$ and smaller $\alpha$ values are associated to a much slower evolution of the momentum distribution.
Thus, the Markov approximation is reached for increasing system sizes, and the larger the range of the interaction (that is,
the smaller $\alpha$), the smaller the system size required to reach it.

\begin{figure}[ht]
\centering
\includegraphics[width=0.5\textwidth]{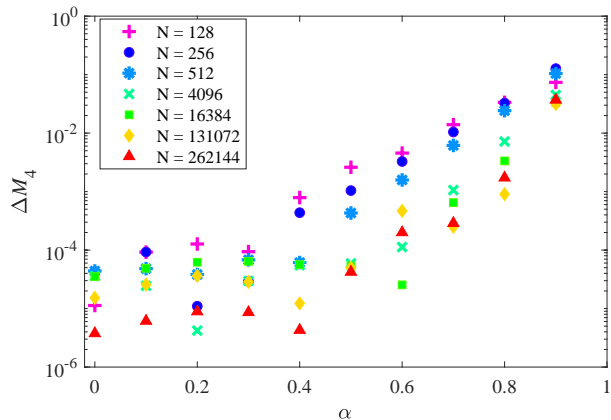} 
	\caption{Relative variation of the fourth momentum $\Delta M_4$, see Eq.(\ref{eq:dm4}), for different system sizes and interaction ranges $\alpha$. The initial value the momentum is always close to $M_4(0)\approx30$. The plain lines correspond to $N=256$, the dash-dotted ones to $N=4096$ and the dashed ones to $N=16384$. Simulations realized for $p_0=7$.}
	\label{Fig:slopes}
\end{figure}

The (non-)Markovian nature of the dynamics has been characterized in details for the mean-field case ($\alpha=0$) in Ref.~\cite{lourenco2015}.
We here simply evaluate the typical relative change of the fourth momentum during the time over which the force auto-correlation cancels,
by characterizing the average growth of the momentum (through a linear fit of $M_4(t)$) as given by the normalized quantity:
\begin{equation}
    \Delta M_4=\frac{\left\langle dM_4/dt\right\rangle t_{FC}}{\left\langle M_4\right\rangle},\label{eq:dm4}
\end{equation}
where $\left\langle \cdots\right\rangle$ stands for a time average between $t=0$ and $t_{FC}$. Its evolution is shown in Fig.~\ref{Fig:slopes}.
Thus, close to the threshold to short-range interactions ($\alpha=1$), the dynamics requires huge system sizes to reach the Markovian regime whereas
close to the infinite-range case ($\alpha=0$), the Markov approximation is already valid for modest system sizes. We note that such results are
highly dependent on the model, its parameters, dimensionality, etc.

\subsection{Non-Markovian regime}

Hence, if over the momentum distribution changes significantly over the time scale during which the force auto-correlation is non-zero, the dynamics must be considered non-Markovian: The lower order term in the right-hand side of Eq.~\eqref{eq12b} does not vanish and contributes significantly to
the single-particle distribution evolution. The timescales of the evolution of the single-particle distribution $f_1$ with time originates in the sum $\sum_k r^{-2\alpha}_{jk}$, whose scaling with the system size changes with the interaction range. Assuming periodic boundary conditions, for which the distance is given by $r_{jk}=\xi\min(|j-k|,N-|j-k|)$, in the large-$N$ limit one obtains
\begin{equation}
        \sum_{k\neq j} \frac{1}{r_{jk}^{2\alpha}} \equiv\frac{1}{\xi^{2\alpha}}\times
        \left\{
                \begin{array}{l}
                        N^{1-2\alpha}\, 2^{2\alpha}/(1-2\alpha),\: {\rm if} \: 0\leq\alpha<1/2;
                        \\ \log N,\: {\rm if} \: \alpha=1/2;
                        \\ \zeta(2\alpha),\:\:{\rm if}\: \alpha>1/2.
                \end{array}
                \right.
        \label{scalingS2}
\end{equation}
Inserting the $1/\tilde{N}^2\sim N^{2\alpha-2}$ in the above expression leads to the following scaling for the relaxation time, as due to the non-Markovian contribution:
\begin{equation}
        \tau_r\sim
        \left\{
                \begin{array}{l}
                        N,\: {\rm if} \: 0\leq\alpha<1/2;
                        \\ N/\log N,\: {\rm if} \: \alpha=1/2;
                        \\ N^{2-2\alpha},\:\:{\rm if}\: \alpha>1/2.
                \end{array}
                \right.
        \label{eq:scalingtau}
\end{equation}
The threshold at $\alpha=d/2$ reported in Ref.~\cite{bachelard2013} is thus already present in the non-Markovian contribution, from two-particle collisions
term (lower order term). This threshold is also clearly visible in the variation of the momentum distribution presented in Fig.~\ref{Fig:slopes}.

\subsection{Markovian regime}

Prediction (\ref{eq:scalingtau}) fits only partially the numerical findings for the one-dimensional classical chain reported in  Ref.~\cite{bachelard2013},
where a scaling of the order of $\tau\sim N^{1.5}$ was observed for $\alpha<0.5$, before it decays roughly as $\tau\sim N^{2.5-2\alpha}$ for $1/2<\alpha<1$.
In other words, the relaxation times appears to scale with a factor of order $\sqrt{N}$ larger than as predicted by the non-Markovian contribution.

The origin of this discrepancy can be found in the competition between non-Markovian and Markovian contributions.
Indeed, as discussed extensively for the mean-field case in Ref.~\cite{lourenco2015}, this competition leads to the
observation of scaling laws that are intermediate between the two regimes: $\tau\sim N$ for the non-Markovian terms
and $\tau\sim N^2$ for the Markovian terms, for $\alpha=0$. The same effect happens for long-range lattices with any $\alpha$,
for the system sizes achievable by numerical simulations.

The inspection of Eqs.~(\ref{eq9}) and~(\ref{eqC32}) reveals that similar $\sum_k r^{2\alpha}_{jk}$ terms appear for the Markovian contribution to the dynamics,
although the variety of terms prevents a thorough analysis of the contribution of $g_3$. Nevertheless, the presence of these sums is an indicator that the
threshold at $\alpha=d/2$ will be preserved at larger system sizes, as supported by the variation of the momentum $M_4$ presented in Fig.~\ref{Fig:slopes}.
There, simulations realized with $N=262144$ (a factor $64$ larger than the relaxation study presented in Ref.~\cite{kastner2011}) still exhibit the threshold,
although such a large system size prevents from realizing a full equilibration study, instead restricting us to the initial stage of the dynamics. Hence, the non-Markovian and Markovian terms present different scaling laws, both exhibiting the $\alpha=d/2$ threshold, and systems of thousands of particles
being in-between these two regimes.

We note that while a rigorous evaluation of the Markovian nature of the dynamics would require a dedicated, and arduous, study of each reduced distribution and auto-correlation function involved in Eq.\eqref{eq9}. This task is beyond the scope of this manuscript, and the good agreement between the prediction of Eq.\eqref{eq:scalingtau} and the numerical results suggest that these terms do not contain scalings that change significantly the (non-)Markovian nature of the dynamics.

\section{Conclusion}

In this work, we have derived a kinetic equation for long-range lattices with power-law interactions. This allowed us to identify contributions that change of scaling with the system size at $\alpha=d/2$, both in the non-Markovian and in the Markovian terms. Lattices of hundreds to thousands of particles are subject to a competition between these two kinds of terms, at least in the context of the homogeneous states of the $\alpha$XY chain considered here.

Establishing a kinetic equations for these classical systems is an important step toward the understanding of their relaxation to equilibrium. The presence of a threshold may lead to a further distinction between the classification of the interaction range, just as it was done between short- and long-range~\cite{Campa2009}, or between lattices and moving bodies~\cite{Gabrielli2010}.

The tools presented here could in principle be generalized to address quantum systems,
by using the quantum analog of the BBGKY hierarchy, e.~g.\ from the Wigner function representation of the
quantum equation of motion~\cite{liboff2003}, and study thermalization processes in quantum systems. In this context,
the emergence of flexible experimental platforms such as cold atom or trapped ion setups,
where the engineering of the Hamiltonians and the reduction of decoherence channels make constant progresses,
is a promising tool to investigate the equilibration processes of both classical and quantum systems.

\begin{acknowledgments}

We thank fruitful discussions with M. Kastner. R. B. benefited from Grants from S\~ao Paulo Research Foundation (FAPESP)
	(Grant Nos. 2014/01491-0 and 2018/01447-2). R.~B. and T.~M.~R.~F benefited from Grants from the National Council
	for Scientific and Technological Development (CNPq) Grant No.\ 	305842/2017-0.

\end{acknowledgments}

\bibliography{refs}

\end{document}